\documentclass{PoS}

\newcommand\apj{{\emph{Astrophysical Journal}}}
\newcommand\apjs{{\emph{Astrophysical Journal, Supplement}}}
\newcommand\aap{{\emph{A\&A}}}
\newcommand\mnras{{\emph{Monthly Notices of the Royal Astronomical Society}}}
\newcommand\pasj{{\emph{PASJ}}}

\title{AGN Population Studies for CTA }

\ShortTitle{AGN Population Studies for CTA }

\author{Yoshiyuki Inoue for the CTA Consortium\\
       Department of Astronomy, Kyoto University, Kitashirakawa, Sakyo-ku, Kyoto 606-8502, Japan\\
        E-mail: \email{yinoue@kusastro.kyoto-u.ac.jp}}

\abstract{Following the great success of the current Imaging Atmospheric Cherenkov Telescopes, the next generation gamma-ray telescope arrays, Cherenkov Telescope Array (CTA), is being prepared. CTA will have an order of magnitude higher sensitivity and wider energy range than current instruments. Large samples of very high energy (VHE; $>30$ GeV) sources will be obtained with CTA. Here we discuss potential AGN population studies for the CTA era concerning the overall source counts and the statistics of high redshift sources. Based on our latest blazar gamma-ray luminosity function (GLF) model, we find that CTA will detect $\sim50$ and $\sim160$ blazars with 1 year and 10 years of blank field sky survey, respectively. CTA is also expected to find a blazar at $z\sim1.4$ ($\sim20$ blazars above $z=1$) based on the blazar GLF. Furthermore, we also examine the detectability of high redshift {\it Fermi} blazars. By extrapolating the {\it Fermi} blazars' spectra with a power-law, we find that CTA has a capability to detect a {\it Fermi} blazar at $z=2.49$.}

\FullConference{AGN Physics in the CTA Era - AGN2011,\\
		May 16-17, 2011\\
		Toulouse, France}

\begin{document}

\section{Introduction}
	Following the great success of the current Imaging Atmospheric Cherenkov Telescopes (IACTs), the next generation gamma-ray telescope arrays, Cherenkov Telescope Array (CTA), is being prepared \cite{CTA}. It is designed to have an order of magnitude higher sensitivity and wider energy range compared to arrays. 
	
	Current IACTs have already found $\sim$100 very high energy (VHE) sources, including more than 40 blazars up to redshift $z=0.536$\footnote{{http://www.mpp.mpg.de/\%7Erwagner/sources/}\\ \ \ \ \ \ \ \ \ \  {http://tevcat.uchicago.edu/}}. Blazars, a class of active galactic nuclei (AGNs), are the dominant population in the extragalactic gamma-ray sky. Almost all of the extragalactic sources detected by EGRET and {\it Fermi} are blazars \cite{har99,abd10_catalog}. In the CTA era, it is naturally expected that the number of blazars detected in the VHE range increases dramatically and that a much deeper VHE universe will be explored. Therefore, it would be possible to carry out a population study of VHE blazars with CTA, which would provide a crucial key to understanding AGN populations, high-energy phenomena around supermassive black holes in AGNs, the cosmological evolution of AGNs, and the distribution of the Extragalactic background light (EBL).
	
	The purpose of this paper is to consider the prospect of future blazar surveys by CTA. For this purpose, the blazar gamma-ray luminosity function (GLF) and spectral energy distribution (SED) are needed. The blazar GLF has been studied in detail in many papers (see e.g. \cite{it09,itm10}). A new blazar GLFs \cite{it09,itm10} has recently developed on the basis of the latest determination of the X-ray luminosity function of AGNs \cite{ued03,has05} featuring a so-called luminosity density evolution. Another new feature of these GLFs is taking into account the blazar SED sequence. The blazar sequence is a feature seen in the mean SED of blazars that the synchrotron and inverse Compton (IC) peak photon energies decrease as the bolometric luminosity increases \cite{fos98,kub98}.  The key parameters in GLF have been carefully determined to match the observed flux and redshift distribution of EGRET blazars by a likelihood analysis. Predictions for the {\it Fermi} mission are in good agreement with the recent {\it Fermi} observed data \cite{itm10,ino11}.  
	
	We utilize the model by \cite{itm10} to predict the expected number and distributions of physical quantities of VHE blazars in a future CTA sky   survey. We also discuss the expected high redshift blazar studies by CTA based on the first year {\it Fermi} source catalog. Throughout this paper, we adopt the standard cosmological parameters of $(h, \Omega_M , \Omega_\Lambda) = (0.7, 0.3, 0.7)$.
	
\section{VHE Blazar Survey with CTA}

There are various survey modes for VHE blazars such as a blank field sky survey and a follow-up survey for selected targets at other wavelengths. Here we mainly consider the case of a blank field sky survey mode following \cite{itm10}. It enables us to find VHE sources without any biases from other wavelength observations. It should be noted that there is already a class of blazars which is not detected by {\it Fermi} but by current IACTs such as 1ES 0229+220 \cite{abd09_tev}. In the CTA era, the number of such sources is expected to increase. This kind of sources will be a new key to understanding the VHE gamma-ray emission mechanism in AGN jets, since the deabsorbed spectrum appear unusually hard which poses challenges to conventional acceleration and emission models (see e.g. \cite{lef11}). Furthermore, they will make a constraint on the intergalactic magnetic field strength by the limits to the expected cascade flux in the GeV energy range \cite{ner10}. Thus, it is important to carry out an unbiased survey with CTA. The case of following up {\it Fermi} blazars is also presented in \cite{itm10} in detail.

\begin{figure}
\centering
\includegraphics[width=120mm]{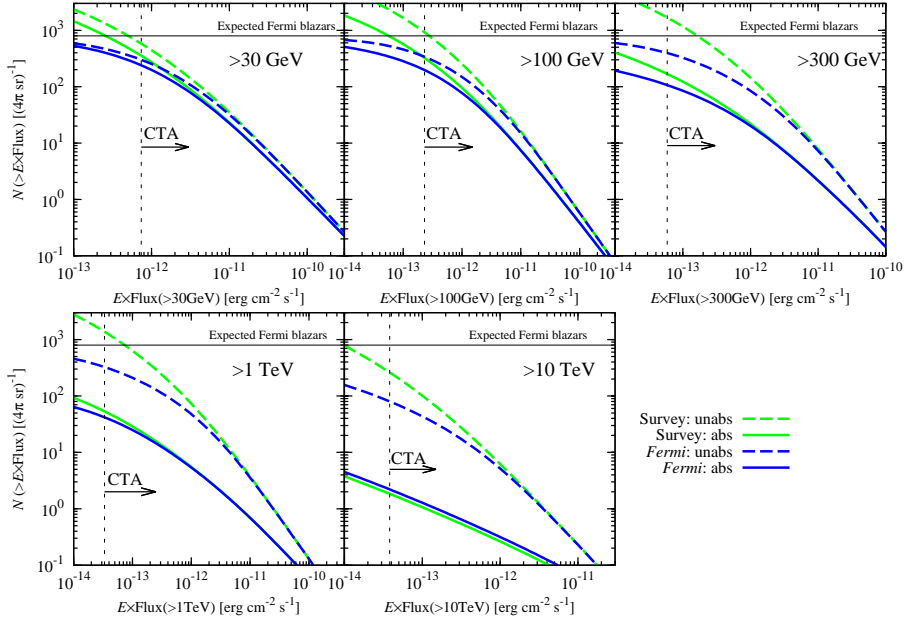}
\caption{	Expected cumulative source counts as a function of the integral gamma-ray flux of VHE blazars. The five panels correspond to different photon energies, as indicated in the panels. The green curves are in the case of blank field sky survey; the blue curves are in the case of following up {\it Fermi} blazars with a sensitivity limit of $3\times10^{-9} $ photons cm$^{-2}$ s$^{-1}$ for photon flux above 100 MeV \cite{atw09}. The intergalactic absorption by EBL is taken into account for the solid curves, but not in the dashed curves. The 5-$\sigma$ detection limits of CTA for the array configuration E and 50 hr observing time  are also shown. The blue solid curve in the panel of 10 TeV is shifted upward artificially by a factor of 1.2 for the purpose of presentation, because the blue solid and green solid curves totally overlap with each other. The horizontal thin solid line is the total expected number of {\it Fermi} blazars with the {\it Fermi} sensitivity given above.}
\label{fig:count}
\end{figure}

To evaluate the expected outcomes of a survey with CTA, we need a blazar SED model and a GLF model. We utilize the recent GLF model by \cite{itm10} which takes into account blazar SED sequence \cite{fos98,kub98}. This GLF model is consistent with the very recent {\it Fermi} observations  \cite{itm10,ino11}. Fig. \ref{fig:count} shows the cumulative source count distributions in the entire sky predicted by \cite{itm10} above the following five energies of 30 GeV, 100 GeV, 300 GeV, 1 TeV, and 10 TeV as indicated in panels. In this figure, we adopt the EBL absorption model by \cite{tot02}, since gamma-ray photons from the universe suffer from the intergalactic absorption effect. The expected number of blazars will not change significantly when we adopt the EBL model \cite{inoinprep}, while it will decrease from our prediction when we adopt the model \cite{kne04}. The differences among EBL models are addressed in \cite{itm10} in detail. We also show the expected counts in the case of following up the {\it Fermi} blazars.

Since observation time is limited, Table \ref{tab:count} shows the expected source counts with 100 hours of total observing time for various survey areas of 40, 400, and 4000 deg$^2$ using the EBL model \cite{tot02}. Since the Field-of-View (FoV) of Large Size Telescope of CTA will be $\sim20$ deg$^2$ \cite{CTA}, we assume FoV of CTA to be 20 deg$^2$. Sensitivities for various observational time per FoV are calculated by using internal CTA tools. CTA will be able to detect $\sim50$ and  $\sim160$ blazars with 1 year and 10 years blank field sky survey assuming 1000 hrs of observing time for 1 year, respectively. The results show that a wide and shallow sky survey is favored for a fixed survey time. A wide field survey is expected to lead to discoveries of serendipitous objects.

\begin{table}[ht]
\begin{center}
\begin{tabular}{cccc}\hline
&50 hr/FoV & 5 hr/FoV&0.5 hr/FoV\\ 
Energy range & 40 deg$^2$ & 400 deg$^2$&4000 deg$^2$\\ \hline
>30 GeV& 0.36 & 1.6 &5.4\\
>100 GeV& 0.35& 1.5& 5.2\\ 
>300 GeV& 0.18& 0.83& 2.8\\ 
>1 TeV& 0.06& 0.25& 0.9\\ 
\hline
\end{tabular}
\end{center}
\caption{Expected blazar source counts for 100 hr blank sky survey by CTA.}
\label{tab:count}
\end{table}

\section{Distant Blazar Study by CTA}

\begin{figure}
\centering
\includegraphics[width=100mm]{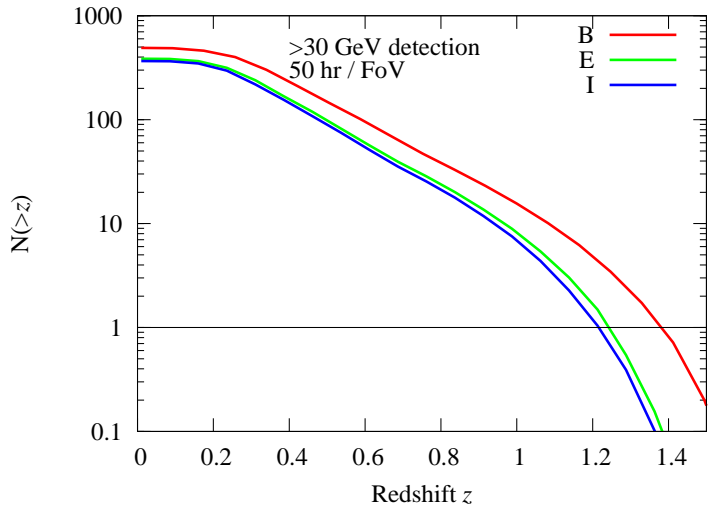}
\caption{Cumulative redshift distribution of blazars above 30 GeV for each CTA array configuration systems (as labeled in the figure), assuming 50 hr observing time for each object and 20 degree zenith angle observation. The horizontal solid line corresponds to an expected count equal to one.}
\label{fig:zcount}
\end{figure}

CTA will also probe a deeper VHE universe than current IACTs do. Fig. \ref{fig:zcount} shows the cumulative redshift distribution of detectable blazars in the entire sky for various CTA array configurations (see \cite{CTA} for details) predicted using the blazar GLF \cite{itm10}. CTA will find $\sim20$ blazars above $z=1$ and the expected highest redshift will be at $z\sim1.4$ using the EBL model \cite{inoinprep}. These blazars will enable us to study the cosmological evolution of VHE AGN populations and to find EBL constraints at high redshift.

\begin{figure}
\centering
\includegraphics[width=100mm]{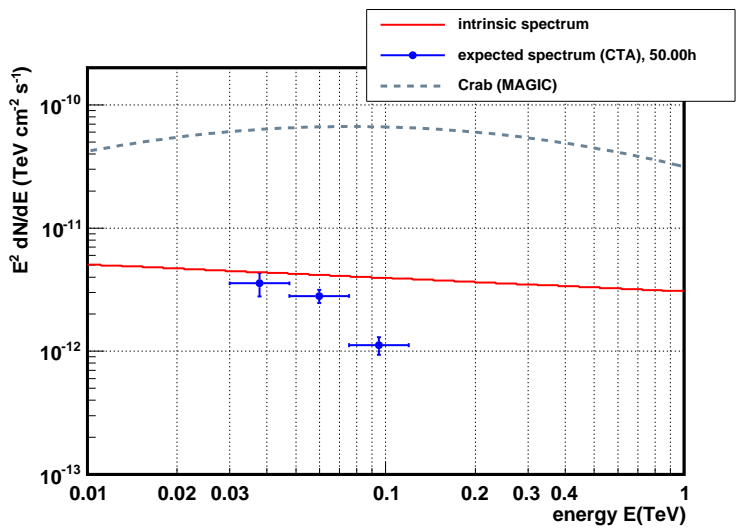}
\caption{Expected spectrum of 1FGL J1344.2-1723  blazar which is located at $z=2.49$, assuming the array configuration B, 50 hr observing time, and 20 degree zenith angle. The red curve shows the intrinsic spectrum without the EBL absorption. The gray curve shows the spectrum of the Crab nebula for comparison. The blue data points show the expected spectral data by CTA including the EBL absorption effects. The expected significance is 9.03-$\sigma$.}
\label{fig:1FGLJ1344-1723}
\end{figure}

Since {\it Fermi} is currently observing the entire GeV sky, we also evaluate the detectability of {\it Fermi} blazars in the first year catalog \cite{abd10_catalog} using the internal spectrum simulation tool and the EBL model \cite{inoinprep}. Fig. \ref{fig:1FGLJ1344-1723} shows the highest redshift {\it Fermi} blazar object at $z=2.49$, 1FGL J1344-1723, detectable by CTA with 50 hr observing time including the intergalactic absorption effect. The expected significance above 30 GeV is 9.03--$\sigma$ with array B. However, we assume a simple power-law extrapolation from the {\it Fermi} spectrum. Further detailed SED fitting studies are required to validate this detectability. For other sources at high redshift, there are 17 {\it Fermi} blazar candidates for CTA above $z=1$ with 50 hr observing time for each sources.

\section{Discussion and Summary}
In this paper, we discussed the expected number of blazars and highest redshift accessible with future CTA observation based on a recent blazar GLF model.  For a blank field sky survey, a wide and shallow search will enable us to carry out an efficient survey. CTA will detect $\sim50$ and $\sim160$ blazars with 1 year and 10 years blank field sky survey, respectively. With CTA, we should be able to find a blazar at $z\sim1.4$ ($\sim20$ blazars above $z=1$). We also estimated the detectable {\it Fermi} blazars. By simply extrapolating the GeV spectrum of the {\it Fermi} blazars in the first catalog, CTA is expected to detect a blazar at $z=2.49$. This will enable us to study the cosmological evolution of VHE blazars.

Although we discussed only blazars in this paper, other classes of AGNs  are also expected as potential sources for CTA, such as core emission from radio galaxies \cite{ino11b}, kpc jet emission \cite{har11}, low luminous AGNs \cite{tak11}. By observing various AGN populations and high redshift sources, CTA will help to unify AGN populations and study the cosmological evolution of the VHE universe.

\section*{Acknowledgments}
The author thanks the MC group in the CTA collaboration for providing the internal MC tool and D. Mazin for providing a tool for evaluating the expected spectrum by CTA. YI acknowledges support by the Research Fellowship of the Japan Society for the Promotion of Science (JSPS).

\end{document}